
\magnification=1200
\baselineskip=20pt
\def\lsim{<\kern-2.5ex\lower0.85ex\hbox{$\sim$}\ }
\def\rsim{>\kern-2.5ex\lower0.85ex\hbox{$\sim$}\ }
\overfullrule=0pt
\ \ \
\vskip 3cm
\centerline{\bf Comment on \lq\lq Disassembling Anyons"}
\vskip 1cm
\centerline{by}
\vskip 1cm
\centerline{C. R. Hagen}
\centerline{Department of Physics and Astronomy}
\centerline{University of Rochester}
\centerline{Rochester, NY 14627}
\vfil\eject

In a recent work$^1$ it has been suggested that a Chern-Simons (CS) theory
with two gauge fields could be profitably invoked in various condensed
matter applications.  By way of reminder it should be pointed out that the
pure CS gauge field theory (as opposed to QED with an auxiliary CS term) was
introduced by the author some years ago$^2$ and was shown to provide the
first concrete realization of anomalous spin in a canonical field theory.
Doubled (and higher multiplicity) CS theories have also seen a number of
applications.  They have, for example, been used to derive the most general
gauge theory in 2 + 1 dimensions$^3$ as well as the most general parity
invariant CS theory.$^4$

In the context of ref. 1 it was important (in order to maximize possible
application to condensed matter problems) to identify this gauge field
doubling with the introduction of spin-up and spin-down electrons.  Although
one can certainly do this in the CS field theory, it is not possible to
discuss the resulting theory (as done in ref. 1) in terms of fractional
statistics and anyons.  Plainly stated, the CS field theory of interacting
spin-${1 \over 2}$ particles does not have an anyonic interpretation.$^4$
 The corresponding (nonrelativistic) spin zero theory \underbar{can} be
reconciled to such a framework and indeed the second virial coefficient of
a gas of such particles clearly displays the periodicity required in the
anyon approach$^5$ (i.e., it is invariant under translation of the flux
parameter $\alpha$ by an integer multiple of two units).  However,
 in the spin-${1 \over 2}$ case it is
known both for the single spin component gas$^4$ as well as for the case of
an equal admixture of both spin components$^6$ that the second virial
coefficient violates the periodicity condition required for an anyon
formulation.  In fact there exists in the literature no attempt to account
for the results of refs. 4 and 6 in terms of anyons despite the fact that
only nonrelativistic quantum mechanics should be required for this problem.

It is clear that this point bears strongly on the proposed application in
ref. 1 to work of Anderson.$^7$  The scattering amplitude of the latter is
compared in ref. 1 to the anyon scattering amplitude which is given as
$$V_{k,k^\prime} \sim {k \times k^\prime \over \vert
 k - k^\prime \vert^2} \eqno(1)$$
for momenta $k, k^\prime$.  Since, however, Eq. (1) follows from the spin
zero CS theory rather than the spin-${1 \over 2}$ CS theory, one should
clearly calculate this quantity by an evaluation of the matrix elements of
$\gamma^\mu D_{\mu \nu} (k - k^\prime) \gamma^\nu$ where the $\gamma$
matrices are as given in ref. 8 and the gauge field propagator in ref. 2.
At nonrelativistic energies this gives an amplitude
$$V_{k, k^\prime} \sim {k \times k^\prime \over \vert k - k^\prime
\vert^2} - {1 \over 4}\ i (s + s^\prime)$$
where $s, \ s^\prime = \pm 1$ are the spin projections of the electrons.
 Thus the scattering of two up or two down electrons has an important
additional contribution in the CS spin-${1 \over 2}$ formulation which is
absent from the anyon result (1).  Whether this furthers or hinders
potential application to condensed matter work is not immediately clear.
The crucial observation is that application of CS field theory to such
matters necessarily entails spin complications which are inimical to the
anyon view.

This work is supported in part by the U.S. Department of Energy Grant no.
 DE-FG02-91ER40685.

\bigskip
\bigskip
\noindent {\bf References}
\medskip
\item{1.} F. Wilczek, Phys. Rev. Lett. {\bf 69}, 132 (1992).
\item{2.} C. R. Hagen, Ann. Phys. (N.Y.) {\bf 157}, 342 (1984).
\item{3.} C. R. Hagen, Phys. Rev. Lett. {\bf 58}, 1074 (1987).
\item{4.} C. R. Hagen, Phys. Rev. Lett. {\bf 68}, 3821 (1992).
\item{5.} D. P. Arovas, P. Schrieffer, F. Wilczek, and A. Zee, Nucl. Phys.
{\bf B251}, 117 (1985).
\item{6.} T. Blum, C. R. Hagen, and S. Ramaswamy, Phys. Rev. Lett.
{\bf 64}, 709 (1990).
\item{7.} P. W. Anderson, Phys. Rev. Lett. {\bf 66}, 3226 (1991).
\item{8.} C. R. Hagen, Phys. Rev. Lett. {\bf 64}, 2015 (1990).
\bye